\documentclass[prl,aps,twocolumn]{revtex4}
\usepackage{epsfig}
\usepackage{bm}
\newcommand{\B}[1]{{\bm{#1}}}

\usepackage[latin1]{inputenc}
\newcommand{\beq}{\begin{equation}}
\newcommand{\eeq}{\end{equation}}
\newcommand{\bea}{\begin{eqnarray}}
\newcommand{\eea}{\end{eqnarray}}
\begin{document}

\title{Oscillatory Instability in Two-Dimensional Dynamic Fracture}

\author{Eran Bouchbinder and Itamar Procaccia}
\affiliation{Dept. of Chemical Physics, The Weizmann Institute
of Science, Rehovot 76100 Israel}

\begin{abstract}
The stability of a rapid dynamic crack in a two dimensional infinite
strip is studied in the framework of Linear Elasticity Fracture
Mechanics supplemented with a modified principle of local symmetry.
It is predicted that a single crack becomes unstable by a finite
wavelength oscillatory mode at a velocity $v_c$,
$0.8c_R<v_c<0.85c_R$, where $c_R$ is the Rayleigh wave speed. The
relevance of this theoretical calculation to the oscillatory
instability reported in the companion experimental Letter is
discussed.
\end{abstract}
\maketitle

{\bf Introduction:} High precision experiments on dynamic fracture
in slabs of amorphous materials revealed very interesting
instabilities in the form of micro-branching. Above a critical
velocity of about $0.4c_R$, where $c_R$ is the Rayleigh wave speed,
a single, straight, rapidly moving crack is unstable against the
appearance of small diameter side-branches that affect both the
morphology and the velocity of propagation \cite{99SF}. An important
observation regarding this instability is that although experiments
are typically performed on quasi-two-dimensional samples (i.e.
samples for which the third dimension is significantly smaller than
the other two dimensions), the instability is intrinsically
three-dimensional \cite{99FM, 05BP};  at the onset of instability
the micro-branches occupy only a small fraction of the small third
dimension, barely misting the mirror quality of the main crack.

Nevertheless, theoretically there were a number of attempts to explain this instability in the context
of Linear Elasticity Fracture Mechanics (LEFM) in two dimensions, where the experimentally observed instability
was interpreted as a macroscopic crack bifurcation. It was shown that above some critical velocity such a bifurcation is allowed on energetic grounds \cite{05AB},
 but not necessarily realized dynamically. The aim of this Letter is to show that two-dimensional LEFM predicts the existence of
 a {\em dynamical} oscillatory instability. We refer the reader to the experimental companion paper where a similar conclusion is
 offered on the basis of crack propagation
in thin films loaded with a fixed grip \cite{Ariel}. Here we construct a
theoretical analog:  a semi-infinite straight crack propagating at a
constant velocity in an infinitely long two-dimensional strip under
fixed grip boundary conditions. The standard framework of LEFM is
supplemented with a modified principle of local symmetry \cite{93HS,
03BHP}, and see below for details. The analysis is based on a
recently derived solution for the linear perturbation problem of the
dynamic stress intensity factors using the weight functions method
\cite{95WM, 97WM, 05MMW}. We find that an oscillatory mode of finite
wavelength becomes unstable above a critical velocity $v_c$, $0.8c_R<v_c<0.85c_R$.

{\bf The perturbation problem:} Consider a semi-infinite straight
crack dynamically propagating at a constant velocity $v$ in an
infinitely long two-dimensional strip of width $2h$ under fixed-grip
boundary conditions. A coordinate system $(x,y)$ is located on the
central line of the strip with $x$ being the direction of crack
propagation. The unperturbed crack configuration $M_0$ at any time
$t$ is described by
\begin{equation}
M_0=\{(x_0,y_0): -\infty < x_0 < vt,~ y_0=0\} \ .
\label{unperturbed}
\end{equation}
We then consider a configuration $M_{\epsilon}$ that results from a
small time-dependent perturbation of the crack path
\begin{equation}
M_{\rm \epsilon}=\{(x_{\rm \epsilon},y_{\rm \epsilon}): -\infty <
x_{\rm \epsilon} < vt+\epsilon\varphi(t),~ y_{\rm \epsilon}=\epsilon
\psi(x)\} \label{perturbed} \ ,
\end{equation}
where $\varphi(t)$ and $\psi(x)$ are smooth dimensionless functions
that define the longitudinal and transverse perturbations,
respectively. The dimensional amplitude $\epsilon$ satisfies $0 <
\epsilon/h \ll 1$, ensuring that both the speed and the path are only slightly
perturbed. For convenience we define a co-moving frame by
the transformation $X=x-vt$ and $Y=y$. The material is assumed to be
linear elastic and isotropic, with the displacement field ${\bf
u}(x,y)$ satisfying Navier's equation
\begin{equation}
(\lambda+\mu)\nabla\left(\nabla\cdot{\bf u}\right)+\mu\nabla^2{\bf
u}~=~\rho\ddot{\bf u} \ , \label{Navier}
\end{equation}
where $\rho$ is the density and the dots stand for partial time
derivatives. $\lambda$ and $\mu$ are the Lam\'e coefficients that
relate the spatial derivatives of ${\bf u}$ to the components of the
stress tensor ${\B \sigma}$ \cite{86LL}
\begin{equation}
\sigma_{ij}=\lambda \delta_{ij} \sum_k \partial_k u_k  + \mu
\left(\partial_i u_j+\partial_j u_i \right) \ , \label{Hooke}
\end{equation}
with $i,j=x,y$. The boundary conditions are given by
\begin{equation}
\sigma_{ij}n_j=0\quad\hbox{on the crack,}\quad \B u(x,\pm h)=\pm w
\B y\ , \label{BC}
\end{equation}
where ${\B n}$ is an outward unit vector normal to the crack faces,
$\B y$ is a unit vector in the y direction and $w$ is a constant. As
the fracture process is localized near the crack tip we are mainly
interested in the fields in that region. To that aim we define a
shifted and rotated frame $(X',Y')$ according to
\begin{equation}
X'+iY'=\{X-\epsilon\varphi(t)+i\left(Y-\epsilon \psi(vt) \right)
\}e^{i\theta} \ , \label{rotate}
\end{equation}
where $\theta$ is the angle between the crack tip orientation and
the X-axis such that
$\tan(\theta)=\epsilon\partial_x\psi(\epsilon\varphi(t)+Vt)$. Then,
the asymptotic expansion of the stress field along the $Y'=0$
direction, sufficiently close to the crack tip ($X'\to 0^{\rm +}$)
yields \cite{98Fre}
\begin{equation}
\sigma_{_{\rm Y'Y'}}(X',0,t) \simeq \frac{K_{_{\rm
I}}(v,t)}{\sqrt{2\pi X'}}, ~~ \sigma_{_{\rm X'Y'}}(X',0,t) \simeq
\frac{K_{_{\rm II}}(v,t)}{\sqrt{ 2\pi X'}} \ , \label{SIFs1}
\end{equation}
where $K_{_{\rm I}}$ and $K_{_{\rm II}}$ are the mode I (tensile)
and II (shear) ``stress intensity factors'' respectively. Our goal
is to obtain $K_{_{\rm I}}$ and $K_{_{\rm II}}$
as functionals of the perturbation functions $\varphi(t)$ and
$\psi(x)$ to linear order in $\epsilon$.

{\bf Solution:} As a result of the long range nature of elastic
interactions we expect $K^{(1)}_{_{\rm I}}$ and $K^{(1)}_{_{\rm
II}}$ to depend on the whole zeroth order solution. Therefore, we
write the displacement field as $\B u\!=\!\B u^{(\rm
0)}\!+\! \B u^{(1)}$ and the stress field as $\B \sigma=\B
\sigma^{(\rm 0)}+ \B \sigma^{(1)}$. Then, we represent the
unperturbed displacement field as $\B u^{(0)}\!=\!\B u^{(\rm
s)}\!+\!\B y wy/h $. Here $\B u^{(\rm s)}$ is a field satisfying Eq.
(\ref{Navier}) in a strip with a straight crack and the boundary
conditions $\sigma_{xy}\!=\!0$, $\sigma_{yy}\!=\!-(2\mu+\lambda)w/h$
on the crack and $\B u^{(\rm s)}(x,\pm h)=0$. In addition, the
asymptotic expansion of the unperturbed stress field is given by
\begin{equation}
\sigma^{(0)}_{_{\rm YY}}(X,0) \simeq \frac{K^{(0)}_{_{\rm
I}}(v)}{\sqrt{2\pi X}}+ A^{(0)}_{_{\rm I}}(v)\sqrt{
 X} \ , \label{SIF0}
\end{equation}
where we have included the sub-leading contribution proportional to
$A^{(0)}_{_{\rm I}}$. Accordingly, the expansion for the dynamic stress intensity
factors is written as
\begin{equation}
K_{_{\rm I}}(v,t) \simeq K^{(0)}_{_{\rm I}}(v)+ K^{(1)}_{_{\rm
I}}(v,t),~ K_{_{\rm II}}(v,t) \simeq K^{(1)}_{_{\rm II}}(v,t) \ .
\label{SIF_expansion}
\end{equation}
Note that there is no zeroth order contribution to $K_{_{\rm II}}$
since the unperturbed crack is pure mode I.

The zeroth order problem is rather straightforward and the
resulting solution is presented in \cite{05MMW}. The goal now is to
obtain $K^{(1)}_{_{\rm I}}$ and $K^{(1)}_{_{\rm II}}$ as linear
functionals of the perturbation functions $\varphi(t)$ and
$\psi(x)$. The crucial step is the construction of an auxiliary
field $\B U$, whose components are the so-called dynamic weight
functions \cite{95WM}. $\B U$ satisfies Eq. (\ref{Navier}) in a
strip with a moving straight crack along the X-axis and possess the
following properties: (i) $\left[\B
U\right](X\!<\!0,0)=0\quad\hbox{and}\quad\B U(X,\pm h)=0$. (ii) The
components of the stress tensor $\sigma_{_{\rm iY}}\left(\B
U\right)$ are continuous and $\sigma_{_{\rm iY}}\left(\B
U\right)(X\!>\!0,0)=0$. (iii) $\left[U_i\right](X,0) \simeq
c_iX^{-1/2}\delta(t)$ for $X \to 0^+$. Here $c_i$ are constants and
$[\cdot]$ denotes the jump of a function across $Y=0$. The field $\B
U$ is found by solving the associated Wiener-Hopf type equations and
the explicit Fourier space representation of $[\B U]$ is given in
\cite{05MMW}. With these objects at hand the original reasoning of
\cite{95WM, 97WM} applies without change and the linear functional
$K^{(1)}_{_{\rm I}}$ and $K^{(1)}_{_{\rm II}}$ are shown to be (see
Ref. \cite{05MMW} for details)
\begin{eqnarray}
K^{(1)}_{_{\rm I}}(v,t)&=&\epsilon K^{(0)}_{_{\rm I}}\{{\cal
Q}_{_{\rm Y}}(t)\ast\varphi(t) \}+\epsilon
\sqrt{\case{\pi}{2}}A^{(0)}_{_{\rm I}}\varphi(t)\\ \label{in-plane}
K^{(1)}_{_{\rm II}}(v,t)&=&\epsilon
\sqrt{\case{\pi}{2}}\psi(vt)\Theta(v)A^{(0)}_{_{\rm
I}}+ \epsilon \partial_x\psi(vt)\Upsilon(v)K^{(0)}_{_{\rm I}} \nonumber\\
&+& \epsilon \Theta(v)\{{\cal Q}_{_{\rm X}}(t)\ast\psi(vt)
\}K^{(0)}_{_{\rm I}} \nonumber\\
&+& \epsilon \{[U_{_{\rm X}}] \ast \langle P^{(1)}_{_{\rm X}}
\rangle - \langle U_{_{\rm Y}} \rangle \ast [P^{(1)}_{_{\rm
Y}}]\}|_{X=0} \ , \label{out-of-plane}
\end{eqnarray}
where $\langle\cdot\rangle$ denotes the average of a function across
$Y=0$ and $\ast$ denotes a convolution with respect to all relevant
variables. The function $\langle U_{_{\rm Y}} \rangle$ can be
expressed in terms of the components of $[\B U]$, while the Fourier
transform of the field ${\cal \B  Q}$ is obtained by a two-term
asymptotic representation of the Fourier transform of $[\B U]$, see
\cite{05MMW}. The functions $\Theta(v)$ and $\Upsilon(v)$ are known
universal function of $v$ \cite{05MMW}; $[P^{(1)}_{_{\rm Y}}]$ and
$\langle P^{(1)}_{_{\rm X}} \rangle$ are given by
\begin{eqnarray}
[P^{(1)}_{_{\rm Y}}] &=& \left(\psi(x)\rho v^2[\partial^2_{xx} u^{(\rm s)}_{_{\rm Y}}](X) \right)H(-X) \ , \nonumber\\
 \langle P^{(1)}_{_{\rm X}}
\rangle &=& \{ \psi(x)T^{(1)}(X)-\partial_x\left(\psi(x)T^{(0)}(X)
\right)\}H(-X), \nonumber\\
T^{(0)}(X)&=&\sigma^{(\rm s)}_{_{\rm YY}}(X,0)+\case{\lambda w}{h},
\nonumber\\
T^{(1)}(X)&=&\rho v^2\partial^2_{xx} u^{(\rm s)}_{_{\rm X}}(X,0) \ .
\end{eqnarray}
Here $H(\cdot)$ is the Heaviside function and recall that
$x=X\!+\!vt$.

The various terms on the RHS of Eq. (\ref{out-of-plane}) have clear
physical meanings. The first term is a result of shifting the crack
tip out of the symmetry line $y=0$ to $y=\psi(vt)$ in the presence
of the unperturbed asymptotic field of Eq. (\ref{SIF0}). By
dimensional analysis it is obvious that this term must be
proportional to the sub-leading contribution $A^{(0)}_{_{\rm I}}$.
The second term on the RHS of Eq. (\ref{out-of-plane}) is a result
of changing the crack tip orientation by an angle $\theta \simeq
\epsilon \partial_x \psi(vt)$ (see Eq. (\ref{rotate})) in the
presence of the unperturbed asymptotic field of Eq. (\ref{SIF0}).
These two terms are both local and instantaneous. As a result of
long range elastic interactions and wave propagation we expect also
non-local contributions in time and space. The third term on the RHS
of Eq. (\ref{out-of-plane}) is a result of the interaction of the
tip with waves that were transmitted by itself at earlier times. The
precise nature of the interaction is carried by the time-dependent
kernel ${\cal Q}_{_{\rm X}}(t)$. The last term on the RHS of Eq.
(\ref{out-of-plane}) represents the interaction of the tip with all
the history of propagation of the crack. This interaction is
non-local both in time and space. The functions $[P^{(1)}_{_{\rm
Y}}]$ and $\langle P^{(1)}_{_{\rm X}} \rangle$ are the effective
mode II (shear) loads on a straight crack, loads that are induced by
the broken up-down symmetry due to $\psi(x)$. With this
interpretation in mind, one observes that the dynamic weight
functions $\B U$ are response functions that quantify the effect of
forces applied to the crack line on the asymptotic stress field near
the tip. An instability, if exists, should emerge as result of the
competition between these various terms.

{\bf Stability analysis:} In order to study the stability of the
straight crack against small perturbations, one must have an
equation of motion for the crack tip. Such an equation does not
emerge from LEFM. This theory can determine the stress and
displacement fields, under the linear stress-strain assumption of
Eq. (\ref{Hooke}), {\em given} the crack and the loading histories,
but cannot tell where and at what rate the crack tip will propagate.
The rate of crack propagation is determined by assuming that the
energy flowing to the crack tip per unit crack extension per unit
time, $J$, equals
$v\Gamma(v)$, where $\Gamma(v)$ is the ``fracture energy''.
The quantity $J$ is calculated using the dynamic stress intensify factors in
the asymptotic expansion Eq. (\ref{SIFs1}) \cite{98Fre}.
Considering a pure longitudinal perturbation, $\psi(x)\!=\!0$ and
$\varphi(t)\!\ne\!0$, one obtains, to first order in $\epsilon$,
\begin{eqnarray}
K^{(0)}_{_{\rm I}}\{{\cal Q}_{_{\rm Y}}(t)\ast\varphi(t)\}+
\sqrt{\case{\pi}{2}}A^{(0)}_{_{\rm I}}\varphi(t) \nonumber\\
= \dot{\varphi}(t) \left(\frac{\partial_v {\cal F}(v)}{2 {\cal
F}(v)} + \frac{\partial_v\Gamma (v)}{2 \Gamma(v)}\right)
\label{Long_pertu} \ ,
\end{eqnarray}
where Eq. (\ref{in-plane}) was used and ${\cal F}(v)$ is a known
universal function \cite{05MMW}. In \cite{05MMW} the case of
constant fracture energy, $\partial_v\Gamma (v)=0$, was studied.
Since for the fixed-grip strip configuration $J=vG$, where the
``energy release rate'' $G$ is constant, every velocity $v$ is
possible as long as $G=\Gamma$. It was shown in \cite{05MMW} that
the crack is unstable for all the velocities {\bf below} $v \approx
0.6c_R$. On the other hand, real materials have $\partial_v\Gamma
(v)>0$ such that a unique velocity is selected according to
$G=\Gamma(v)$ and longitudinal stability is expected \cite{06BPP}.
We verified that Eq. (\ref{Long_pertu}) indeed predicts
longitudinal stability for all velocities as long as
$\partial_v\Gamma (v)>0$. Accordingly, we can safely assume a
straight crack propagation at a constant velocity and focus on
transverse perturbations, i.e. $\psi(x)\!\ne\!0$ and
$\varphi(t)\!=\!0$. Note that all the calculations, here and
below, were performed for a material with a fixed ratio between
$\lambda$ and $\mu$ such that Poisson's ratio for plane stress
condition was $\nu=0.23$ \cite{86LL}. The results depend smoothly (and weakly)
on $\nu$.
\begin{figure}
\centering \epsfig{width=.52\textwidth,file=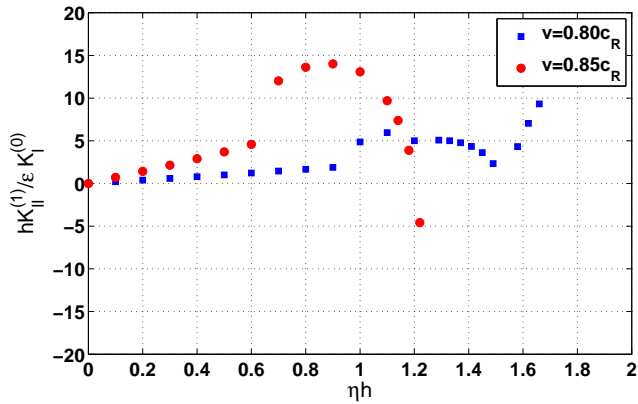}
\caption{(color online) $hK^{(1)}_{_{\rm II}}/\epsilon K^{(0)}_{_{\rm I}}$ as a
function of $\eta h$ for two different velocities. The squares
correspond to $v=0.8 c_R$ and the circles correspond to $v=0.85c_R$.
The onset of instability occurs at $v_c$, $0.8c_R<v_c<0.85c_R$, where
$K^{(1)}_{_{\rm II}}$ first changes sign from positive to negative.
The wave number at the onset of instability is approximately given by $\eta_c h\!
\approx\! 1.4$.} \label{unstable}
\end{figure}

To study the transverse stability of the straight crack we employ the equation
suggested in \cite{93HS}
\begin{equation}
\frac{\partial\theta}{\partial t}=-f(K_{_{\rm I}},K^2_{_{\rm II}},v)
K_{_{\rm II}} \ , \label{HS}
\end{equation}
where $f(K_{_{\rm I}},K^2_{_{\rm II}},v)$ is {\em a positive
definite} dynamic material function that quantifies how the
asymmetry in the asymptotic fields Eq. (\ref{SIFs1}), characterized
by $K_{_{\rm II}}$, is transferred to the crack tip itself; $\theta$
is defined in Eq. (\ref{rotate}). Eq. (\ref{HS}) was applied quite
successfully in \cite{03BHP, 05BMP}. The equation was obtained by
using  symmetry arguments and assuming (i) that the crack tip is not appreciably blunted
and the crack path is a trace of a point,
(ii) that the region around the tip in which linear elasticity theory is not applicable (the so-called ``process zone") is small compared to any other
length scale, (iii) the crack path is smooth. By dimensional considerations
we define a dynamic length scale  as
\begin{equation}
\ell(v) \equiv \frac{v}{f(K_{_{\rm I}},K^2_{_{\rm II}},v) K_{_{\rm
I}}} \ . \label{l_pz}
\end{equation}
As this length scale is expected to be small, in our case implying
$\ell(v)/h \ll 1$, Eq. (\ref{HS}) is approximated by many authors by
$K_{_{\rm II}}/K_{_{\rm I}} \approx 0$, which is the well-known
``principle of local symmetry'' \cite{74GS}.  Eq. (\ref{HS}) is thus
a modified principle of local symmetry.  Note that for $\ell(v)$ to
be well defined in the limit of small $v$, we must have $f(K_{_{\rm
I}},K^2_{_{\rm II}},v) \propto v$ in that limit.

To proceed we can consider two types of transverse perturbations, one localized right at the tip \cite{05MMW}, and the other global, like a small amplitude sine function \cite{95ABP}. We follow
the latter approach since real cracks are never perfectly straight. Analyzing the perturbation
in linear modes, we consider
\begin{equation}
\psi(X+vt)=\sin\{\eta(X+vt)\} \ .\label{perturb}
\end{equation}
Then, we note that Eq. (\ref{HS}), with $\partial_t
\theta \!\simeq\! \epsilon v \partial_{xx}\psi(vt)$, is a linear
integro-differential equation for $\psi(x)$. In principle, such an
equation can be analyzed by standard methods, but we encounter a
difficulty as $f(K_{_{\rm I}},K^2_{_{\rm II}},v)$ is not yet known.
To overcome this difficulty, we focus, without loss of generality, on times such that
$\psi(vt)=0$. The dynamical equation (\ref{HS}) then predicts that
if $\theta$ and $K_{_{\rm II}}$ have the same sign, with $f>0$, then
$\partial_t\theta$ has the opposite sign and $|\theta|$ decreases,
which means that a small perturbation decays. By the same argument,
for $\theta$ and $K_{_{\rm II}}$ having the opposite sign a small
perturbation grows. This criterion is identical to the one used in
Refs. \cite{95ABP, 03BHP}. Therefore, considering the perturbation
introduced in Eq. (\ref{perturb}), we interpret a change of
$K_{_{\rm II}}$ from positive to negative as an instability. Note
that this criterion is independent of the exact form of the function
$f(K_{_{\rm I}},K^2_{_{\rm II}},v)$ and the wavelength of an
unstable mode is expected to be correct up to modifications of the order
of $\ell(v_c)$. Having an explicit instability criterion at hand, we
calculate numerically $K^{(1)}_{_{\rm II}}$ in Eq.
(\ref{out-of-plane}) as a function of the wave number $\eta$ for
various velocities and look for the lowest velocity for which an
intersection with the x-axis occurs. In Fig. \ref{unstable} we show
$hK^{(1)}_{_{\rm II}}/\epsilon K^{(0)}_{_{\rm I}}$ as a function of
$\eta h$ for two representative velocities, one below the
instability and one above. We observe that for $v\!=\!0.8 c_R$ the
curve is always positive, implying stability, while for $v\!=\!0.85
c_R$ the curve changes sign from positive to negative, implying
instability. The normalized wave number $\eta_c h$
at threshold is approximately $\eta_c h\! \approx\! 1.4$. The instability should
be understood as resulting from a competition between local
stabilizing terms and non-local destabilizing terms in Eq.
(\ref{out-of-plane}) (see discussion above), which we believe to be
the dynamic counterpart of the phenomenon studied in \cite{95ABP,
03BHP}.

{\bf Discussion and summary:}  While oscillatory instabilities were observed and discussed before in a number of fracture contexts
(\cite{93YS} in the context of quasi-static thermal cracking, \cite{94ABRR} in the context of lattice simulations, \cite{02DPMS} in large biaxial strain experiment in rubber, \cite{04HL} in context of
a phase field model), it appears that the present is the first calculation predicting such a high velocity
instability, purely on the basis of LEFM. This prediction is in correspondence with the experimental
observations in quasi-two-dimensional thin layers where the side branching instability is suppressed
(see the companion Letter \cite{Ariel}).

Notwithstanding the correspondence of the predicted onset of oscillatory instability compared to the
experiment, one should point out that the predicted wavelength of the saturated post-instability solution
differs qualitatively in theory and experiment. The theory, being based on LEFM  that contains
no intrinsic scales, is bound to predict wavelength that scales with the width $h$ of the strip. The
experiment predicts saturated oscillatory cracks with wavelength that do not depend on the width of the strip. Interestingly enough, the experiment also reports an excellent agreement with LEFM for
all the properties of the dynamic crack up to the instability point. It is tempting to interpret these findings
by stating that LEFM supplemented with the modified principle of local symmetry is trustworthy
up to the instability, but then fails once the oscillatory motion is stabilized. The indication is that while
the crack is straight all the non-elastic aspects can be lumped into an effective fracture energy
$\Gamma(v)$. The true free boundary dynamics of the crack, once deviating from the straight
line, call for a new understanding that must go beyond LEFM with or without principles of local symmetry. It is very possible that the actual shape of the tip, including its radius of curvature, the distortion with respect to the symmetry axes and all other dynamic degrees of freedom become essential
in determining the actual path of the crack. Such degree of freedom can introduce intrinsic length and/or time scales. The only way to reach such a theory is to derive
equations directly for the free boundary that defines the crack. A theory that advances in this
direction will be presented in due course \cite{06BLLP}.

{\bf Acknowledgments}: We thank A. B. Movchan and N.V. Movchan for their indispensable help in the numerical analysis of the instability equations, and J.R. Willis for useful discussions.  Vincent Hakim explained to us the range of validity of the modified principle of local symmetry. This work has been supported in part by the Israel Science Foundation. E. B. is supported by the Horowitz Complexity Science Foundation.


\begin{thebibliography}{99}

\bibitem{99SF} E. Sharon and J. Fineberg, Nature {\bf 397} 333 (1999).

\bibitem{99FM}
J. Fineberg and M. Marder, Phys. Rep. {\bf 313}, 1 (1999).

\bibitem{05BP} E. Bouchbinder and I. Procaccia, Phys. Rev. E {\bf 72}, 055103(R) (2005).

\bibitem{05AB}
M. Adda-Bedia, J. Mech. Phys. Solids {\bf 53}, 227 (2005).

\bibitem{Ariel}
A. Livne, O. Ben-David and J. Fineberg, Submitted to Phys. Rev. Lett, (companion Letter).

\bibitem{93HS}
J. Hodgdon and J. Sethna, Phys. Rev. B {\bf 47}, 4831 (1993).

\bibitem{03BHP} E. Bouchbinder, H. G. E. Hentschel and I. Procaccia, Phys. Rev. E {\bf
68}, 036601 (2003).

\bibitem{95WM} J. R. Willis and A. B. Movchan, J. Mech. Phys. Solids {\bf 43}, 319, 1369 (1995).

\bibitem{97WM} J. R. Willis and A. B. Movchan, J. Mech. Phys. Solids {\bf 45}, 591 (1997).

\bibitem{05MMW}
N. V. Movchan, A. B. Movchan and J. R. Willis, QJMAM {\bf 58}, 333
(2005).

\bibitem{86LL}
L. D. Landau and E. M. Lifshitz, {\em Theory of Elasticity}, 3rd ed.
(Pergamon, London, 1986).


\bibitem{98Fre} L. B. Freund, {\em Dynamic Fracture Mechanics},
(Cambridge, 1998).

\bibitem{06BPP} E. Bouchbinder, A. Pomyalov and I. Procaccia, ``Dissipative Visco-plastic Deformation in Dynamic Fracture:
Tip Blunting and Velocity Selection'', Phys. Rev. Lett, in press.

\bibitem{05BMP}
E. Bouchbinder, J. Mathiesen, I. Procaccia, Phys. Rev. E, {\bf 71}
056118 (2005).

\bibitem{74GS}
R.V. Gol'dstein and R. Salganik, Int. J. Frac. {\bf 10}, 507 (1974).

\bibitem{95ABP}
M. Adda-Bedia and Y. Pomeau, Phys. Rev. E \bf{52}\rm, 4105 (1995).

\bibitem{93YS}
A. Yuse and M. Sano, Nature (London) {\bf 362}, 329 (1993).

\bibitem{94ABRR}
F.F. Abraham, D. Brodbeck, R.A. Rafey adn W.E. Rudge, Phys. Rev. Lett. {\bf 73} 272 (1994).

\bibitem{02DPMS}
R. D. Deegan, P. J. Petersan, M. Marder and H. L. Swinney, Phys. Rev. Lett. {\bf 88}, 014304 (2002).

\bibitem{04HL}
H. Henry and H. Levine, Phys. Rev. Lett. {\bf 93}, 105504 (2004).

\bibitem{06BLLP}
E. Bouchbinder, J.S. Langer, T.S. Lo and I. Procaccia, ``Free-Boundary Dynamics in Elasto-plastic Amorphous Media: The Circular Hole Problem and Boundary Layer Approximations", to be published.













\end{thebibliography}
\end{document}